\documentclass[a4paper,aps,pra,showpacs,preprintnumbers]{revtex4}
\usepackage{graphics,graphicx,epsfig}        
\usepackage[centertags]{amsmath}
\usepackage{amsfonts}
\usepackage{amssymb}                     
\usepackage[english]{babel}         
\usepackage{textcomp}                           
\usepackage{amsthm,color,ulem}          
\usepackage{euscript}

\begin{document}

\def\BY{\begin{eqnarray}}
\def\EY{\end{eqnarray}}
\def\BE{\begin{equation}}
\def\EE{\end{equation}}
\def\BEA{\begin{eqnarray}}
\def\EEA{\end{eqnarray}}
\def\k{\mathbf{k}}
\def\q{\mathbf{q}}
\def\r{\vec{r}}
\def\ro{\vec{\rho}}
\def\L{\label}
\def\nn{\nonumber}
\def\({\left (}
\def\){\right )}
\def\[{\left [}
\def\]{\right]}
\def\<{\langle}
\def\>{\rangle}
\def\h{\hat}
\def\hs{\hat{\sigma}}
\def\td{\tilde}
\def\ds{\displaystyle}


\title{Storage and conversion of quantum-statistical properties of light in the resonant quantum memory on tripod  atomic configuration}
\author{ A.S.~Losev, {K.S.~Tikhonov}, {T.Yu.~Golubeva}, {Yu.M.~Golubev} }
\address{St. Petersburg State University, 198504, St. Petersburg, Petershof, ul. Ulianovskaya 1, Russia}
\date{\today}

\begin{abstract}
We have considered theoretically the feasibility of the broadband quantum memory based on the resonant tripod-type atomic configuration. In this
case, the writing of a signal field is carried out simultaneously into two channels, and characterized by an excitation of two spin waves of the
atomic ensemble. With simultaneous read out from both channels quantum properties of the original signal are mapped on the retrieval pulse no worse
than in the case of memory based on $\Lambda$-type atomic configuration. At the same time new possibilities are opened up for manipulation of quantum
states associated with sequential reading out (and/or sequential writing) of signal pulses. For example, the pulse in squeezed state is converted
into two partially entangled pulses with partially squeezed quadratures. Alternatively, two independent signal pulses with orthogonal squeezed
quadratures can be converted into two entangled pulses.
\end{abstract}
\maketitle


\section{Introduction}

Under an optical quantum memory is most often understood the process of light-matter interaction, which may be divided into three stages: "writing"\
--- the mapping of quantum state of light on the state of matter, "storage"\ of the state and "read out"\, that is the retrieving of light with
quantum state close to the original. Thus, the protocols of quantum memory have one main goal: to store and then to retrieve the quantum state with
high efficiency and fidelity. Since the quantum memory is considered as a resource for quantum information and communication networks, it is no less
important aspect of the multimode quantum memory as possibility to manipulate with a few qubits simultaneously \cite{Walmsley-2013, Jaksch-2008,
Gisin-2007, Gisin-2013, Sorensen-2013, Tittel-2014, Polzik-2010, Golubeva_Giacobino-2011, Saunders-2015, Jobez-2015}. In addition to the passive task
of information storage \cite{Lvovsky-2009, Hammerer-2010}, quantum memory can also be used to convert quantum states directly in the same cell. This
approach requires not one but several (at least two) degrees of freedom of the cell memory, allowing to store different quantum states of light. In
this regard, ensembles with multilevel atoms are interesting. Typically, multilevel factor is considered as deleterious, leading to losses during
storage of signals \cite{Mishina-2011,Mishina-2012}. We demonstrate here that basing on a more complicated atomic configuration and choosing the
proper way to retrieve the signal, we not only do not earn additional losses in the system, but also get the opportunity to transform initial quantum
states straight in the memory cell. In this article we consider the quantum memory based on the four-level tripod atomic configuration as a source of
pairs of entangled light pulses. Classical aspects of interaction of fields with atoms in tripod configuration were discussed in \cite{Losev-2013}.

There are several options for implementation of quantum optical memory which differ in duration of interaction, as well as the configuration and
geometry of the fields involved in the process of storage (e.g. see \cite{Lvovsky-2009}). For us in this work the protocol of "high-speed (broadband)
resonant quantum memory"\ \cite{Reim-2010, Golubeva-2011, Golubeva-2012} will be of the most interest as one of the closest to real requirements.
Here as in many other approaches the $\Lambda$-type atomic ensemble (in appropriate basis of states) is used as a storage system. Due to collective
properties of atomic ensemble the scheme is suitable for the writing of short light pulses even when one suppose that their duration (writing time)
is much shorter then a life time of the excited state. This is because the absorption band of light from the ground state is not determined by the
absorption band $\gamma$ of individual atom but by the band $d\gamma$ of all the collective (where optical depth $d$ can achieve 100 or even higher).

In the framework of high-speed resonant quantum memory protocol we will consider the interaction of the ensemble of atoms with light pulses. Herewith
the read out process of the signal will be performed in two different variants allowing to obtain different (but controllable) quantum states of the
retrieved light.

We believe it is important to demonstrate that an additional noise source (associated with the inclusion of an additional energy sublevel and with a
random distribution of atoms between sublevels in the writing process) does not lead to loss of efficiency and quantum correlations in the system
with adequate choice of the read out procedure.

The article is structured as follows. In Section \ref{II} main equations describing processes of light-matter interaction in tripod configuration are
obtained. In Section \ref{III} a transition to $\Lambda$ atomic configuration and an appropriate conversion of the system of equations are described.
Here we also obtained a solution for spin subsystem during the writing process of the signal pulse in a resonant medium. In Section \ref{IV} dynamic
and quantum-statistical properties of the input signal are defined. We considered the radiation of sub-Poissonian laser as the signal. In Sections
\ref{V} and \ref{VI} quantum-statistical properties of spin waves, which arise due to the mapping of this light into the medium, are discussed. In
Section \ref{VII} different variants for observing of quantum correlations of spin waves are discussed and the necessity to pass to the Schmidt modes
(see Section \ref{VIII}) for the correct analysis of the storage of quantum features is substantiated. Finally, Sections \ref{IX} and \ref{X} are
devoted to the discussion of two variants of signal retrieval from the tripod-type memory and the analysis of quantum-statistical properties of the
output light in depending on the read out process.


\section{Main equations of high-speed resonant memory \L{II}}
In this article we consider the tripod-type atomic ensemble as a storage system. In accordance with Fig.~\ref{pict1}, each atom can be represented by
four actual stationary states. Two driving fields with Rabi frequencies $\Omega_1$ and $\Omega_2$  interact resonantly with optical transitions
$|1\rangle-|4\rangle$ and $|2\rangle-|4\rangle$, respectively. In our consideration we assume that the transitions differ in frequency so that each
of driving field interacts with only one of them. Initially all atoms are pumped into the state  $|3\rangle$. It can be achieved, for example, by the
use of an optical pumping \cite{Sprague-2013}. We will aim at storing the quantum state of the signal field which operates on the transition
$|3\rangle-|4\rangle$ resonantly.

\begin{figure}
\centering
\includegraphics[height=34mm]{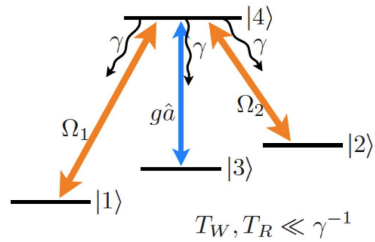}
\caption{Tripod type atomic configuration, $\Omega_1$ and $\Omega_2$ are Rabi frequencies of the driving fields, $\h{a}$ is the slowly varying
amplitude of the signal field.} \L{pict1}
\end{figure}
A thermal motion will be ignored under the assumption that the medium is prepared inside the effective atomic trap. We will not take into account any
relaxation processes supposing the light-matter interaction much faster than the typical times of spontaneous relaxation. We consider three lower
states as long-lived and believe that their excitation is preserved long enough. It provides the complete emptying of the upper state $|4\rangle$
during storage.

These requirements can be satisfied, for example, on a basis of the $D2$-transition of a hyperfine structure of atoms of Rubidium $87: 5^2S_{1/2}
(F=1) \leftrightarrow 5^2P_{3/2} (F=0)$ when a degeneracy of the lower sub-level $5^2S_{1/2} (F=1)$ is lifted by a stationary magnetic field.
Sub-levels $5^2P_{3/2} \; F=0$ and $F=1$ are close to each other ($\cong 0.3 \, \mu eV$, $72.2\,MHz$), therefore one should apply the magnetic field
about $1 G$ \cite{Steck-2010}.

Physical conditions formulated above  are mostly coincide with the conditions in the article \cite{Golubeva-2012}. For that reason the
light-matter interaction Hamiltonian in the dipole approximation obtained there can be easily generalized on the case of tripod configuration:
\BE
\h{V} = i \hbar  \int\limits_{(V)} \!d^3r  [g \h{a}(\r,t) \hat \sigma_{43} (\r,t) e^{i k_s z } + \Omega_1
\hat \sigma_{41}(\r,t) e^{i k_1 z  } +\Omega_2 \hat \sigma_{42}(\r,t) e^{i k_2 z }] + h.c.\L{1}
\EE
Here $k_s$, $k_1$ and $k_2$ are respectively wave numbers of signal and two driving fields. The coupling constant $g$ defines the force of the dipole
interaction of the signal field with a single atom on the transition $|4\rangle-|3\rangle$ with a dipole moment $d_{43}$. It can be written in form
\BE
g = \(\frac{\omega_s}{2\varepsilon_0 \hbar c}\)^{1/2} \! d_{43}.
\EE
The consideration is limited by the rotating wave approximation where all fields are treated as quasi-monochromatic and their carrier frequencies
coincide with the frequencies of atomic transitions: $\omega_s=\omega_{43}$, $\omega_1=\omega_{41}$ and $\omega_2=\omega_{42}$.

The positive-frequency operator of the electric field of a quasiplane and a quasimonochromatic wave travelling in the $+z$ direction can be written
in terms of space- and time-dependent photon annihilation operator $\h{a}(\vec r,t)$ of the signal as:
\BE
\h{E}_s(\vec r,t) = i\(\frac{\hbar \omega_s}{2\epsilon_0 c}\)^{1/2} \! e^{i ( k_s z - \omega_s t)}
\h{a}(z,\vec \rho,t) + h.c.,\qquad \vec r=\(z,\vec\rho\),\qquad\vec\rho=(x,y).
 \EE
Creation and annihilation operators $\h{a}^\dag(\r,t)$ and  $\h{a}(\vec r,t)$ obey the following commutation relations \cite{Kolobov-1999}:
\BEA
&&\[\h{a}(\r,t),\h{a}^\dag(\r^\prime,t)\] = c \; \(1-\frac{i}{k_s} \partial_z - \frac1{2 k^2_s} \Delta_\bot
\) \; \delta^3 (\r - \r^\prime),\qquad\Delta_\bot =
 \partial^2_x + \partial^2_y\\
&&\[\h{a}(z,\ro,t),\h{a}^\dag(z,\ro^\prime,t^\prime)\] = \delta^2 (\ro- \ro^\prime) \;\delta(t - t^\prime),
\EEA
and are normalized so that the mean value $\langle\h{a}^\dag(\vec r,t)\h{a}(\vec r,t)\rangle$ determines the mean number of photons
passing through the cross section per unit time with a dimension of $sec^{-1}cm^{-2}$.

According to \cite{Golubeva-2012}, collective atomic variables (coherences and populations) are introduced as a linear superpositions of
individual variables:
\BE \hs_{i\neq k}(\vec r,t) = \sum_a \hs^a_{i\neq k}(t)\; \delta^3(\vec r - \vec r_a) \quad \mbox{и} \quad
\h{N}_i (\vec r,t) = \sum_a \hs^a_{ii}(t) \;\delta^3(\vec r - \vec r_a),\qquad i,k =1,2,3,4.\L{6}
\EE
Here vector $\vec r_a$ indicates a position of the individual atom with an index $a$. The set of operators $\hs^a_{i k}(t)$ corresponds to individual
atoms, and there are well known relations $\hs^a_{i k}(t)=\hs^a_{i l}(t)\hs^a_{l k}(t)$. Taking into account that variables of different atoms
commute with each other it is easily to get commutation relations for collective atomic variables:
\BE
[\hs_{ik}(\vec r,t), \hs_{ki}(\vec r^\prime,t)] = (\h{N}_i(\vec r,t) - \h{N}_k(\vec r,t)) \;\delta^3(\vec
r - \vec r^\prime),
\EE
Now we have all required to derive a system of differential equations in partial derivatives for variables of the field and the matter in Heisenberg
representation on the basis of Hamiltonian (\ref{1}). We omit details of this procedure, they are described quite well in previous discussions. As a
result, we get the following set of equations:
\BEA
&&\(\frac1c \partial_t + \partial_z - \frac{i}{2 k_s} \Delta_\bot\) \h{a} = - g \hs_{34},\nn \\
&&\partial_t \hs_{12} = - \Omega_1 \hs_{42} \L{HL13} - \Omega_2 \hs_{14},\nn\\
&&\partial_t \hs_{31} = - \Omega_1 \hs_{34} - g \h{a} \hs_{41}, \nn\\
&&\partial_t \hs_{32} = - \Omega_2 \hs_{34} - g \h{a} \hs_{42}, \nn\\
&&\partial_t \hs_{34} = g \h{a} ( \h{N}_3 - \h{N}_4 ) + \Omega_1 \hs_{31} + \Omega_2 \hs_{32}, \L{8.}\\
&&\partial_t \hs_{41} = \Omega_1 ( \h{N}_1 - \h{N}_4 ) + g \h{a}^\dag \hs_{31} + \Omega_2 \hs_{21},\nn \\
&&\partial_t \hs_{42} = \Omega_2 ( \h{N}_2 - \h{N}_4 ) + g \h{a}^\dag \hs_{32} + \Omega_1 \hs_{12} ,\nn\\
&&\partial_t \h{N}_1 = - \Omega_1 ( \hs_{14} + \hs_{41} ),\nn\\
&&\partial_t \h{N}_2 = - \Omega_2 ( \hs_{24} + \hs_{42} ),\nn\\
&&\partial_t \h{N}_3 = - g \h{a}^\dag \hs_{34} - g \h{a} \hs _{43}.\nn
 \EEA
To close this system of equations we have to complete it by the equality $\h{N}_1+ \h{N}_2+\h{N}_3+\h{N}_4=N$, where $N$ is time independent
concentration of atoms involved in the interaction process. Let us remind that according to definitions all of collective variables (and also
concentration $N$) have a fine-grained spatial structure with a characteristic scale of the order of an average distance between immovable atoms.

We omit here well known assumptions \cite{Gorshkov-2007,Golubeva-2012}, which allow us to pass from Eqs. (\ref{8.}) to simplified ones in the form
\BEA
&&\partial_z \h{a} = -g\sqrt N\;\hat c, \L{9}\\
&&\partial_t \hat c = g\sqrt N \;\h{a}+ \Omega_1\hat b_1 + \Omega_2\hat b_2,\\
&&\partial_t \hat b_1 = -\Omega_1 \hat c,\\
&&\partial_t \hat b_2 = -\Omega_2 \hat c.\L{11}
\EEA
Under the deriving of these equations, the averaging on the above-mentioned fine-grained spatial structure is applied. Hereafter we treat $N$ and
other atomic variables as average values.

In set (\ref{9})-(\ref{11}) we introduced renormalized operator amplitudes:
\BE
\hat c = \hat\sigma_{34}/\sqrt N,\qquad\hat b_1 = \hat\sigma_{31}/\sqrt N,\qquad\hat b_2 =
\hat\sigma_{32}/\sqrt N,
\EE
for which canonical commutation relations obey:
\BE
\[\hat c(z,t),\hat c^\dag(z^\prime,t)\]= \delta(z-z^\prime),\qquad\[\hat b_i(z,t),\hat
b_j^\dag(z^\prime,t)\]=\delta_{ij}\delta(z-z^\prime),\qquad i,j=1,2.
\EE
Now the considered problem can be treated as an interaction of four quantum oscillators with Heisenberg amplitudes $\hat a,\;\hat b_1,\;\hat
b_2,\;\hat c$.

Further we will assume that initially all atoms are pumped on the level $|3\rangle$ and will neglect changing of the population of this state caused
by the interaction with actual fields. To satisfy this it is enough to suppose that the number of photons in the signal pulse (which is absorbed on
the transition $|3\> - |4\>$) is much smaller than the population of the level $|3\rangle$.

Also in this article further we will not take into account the transverse structure of the field and the atomic ensemble in particular neglecting the
diffraction of light under its propagation through the medium. We assume that one-dimensional approximation is applicable. It means that actual
fields can be treated as the set of plane waves travelling in the $z$-direction through the atomic layer with thickness $L$. Then in the original
field-matter interaction Hamiltonian the integration over the volume is replaced by one-dimensional integration along the axis $z$ ranging from zero
to the thickness of medium $L$. In equality (\ref{6}) for collective atomic variables, three-dimensional delta-function $\delta^3(\vec r-\vec r_a)$
should be replaced by one-dimensional function $\delta(z-z_a)$. Due to this change the dimensions of collective variables and the concentration $N$
turn out to be equal $cm^{-1}$.


\section{Excitation of spin waves in atomic ensemble \L{III}}
In future we will assume that the Rabi frequencies of both driving fields are equal to each other: $\Omega_1=\Omega_2\equiv\Omega/\sqrt2 $. Let us
introduce instead of spin amplitudes $\hat b_i$ their linear combinations:
\BE
\hat b_\pm=(\hat b_1\pm\hat b_2)/\sqrt2 \L{lc}.
\EE
Then instead of Eqs. (\ref{9})-(\ref{11}) we get
\BEA
&&\partial_z \h{a} = - g \sqrt{N} \h{c}, \L{17}\\
&&\partial_t \h{c} = g \sqrt{N} \h{a} + \Omega \;\h{b}_+, \L{EQc}\\
&&\partial_t \h{b}_+ = -\Omega \h{c}, \L{EQbp}\\
&&\partial_t \h{b}_- =  0. \L{20} \EEA
The first three equations represent the closed system. They are exactly the same as for the three level atoms in $\Lambda$-configurations interacting
with relevant fields \cite{Golubeva-2012}. Let us remind, in that case driving fields are defined by Rabi frequency
$\Omega=\sqrt2\Omega_1=\sqrt2\Omega_2$. Thus we can use the solutions obtained in \cite{Golubeva-2011, Golubeva-2012}, and derive the expression for
the amplitude $\hat b_+$ in the explicit form:
\BE
\hat{  b}_+(  z)= - \frac{1}{\sqrt2} \int^{T_W\!\!\!\!\!\!}_0 dt \; \h{  a}_{in}(  T_W - t) G_{ab}(  z,
t) + \hat v_+( z, T_W),\qquad\hat{  b}_+(  z)\equiv\hat{  b}_+(  z,  T_W).\L{21}
\EE
This formula represents the quantum state of the medium at the end of the writing process $t=T_W$. The kernel $G_{ab}(z,t)$ has a following form:
\BE
G_{ab}(z,t)= \int^{t\!\!\!\!\!\!}_0 dt^\prime e^{-i  t^\prime } \; J_0 \(\sqrt{ z t^\prime}\)\Theta^W
(t^\prime) \;e^{i (t - t^\prime)} J_0 \(\sqrt{z(t - t^\prime)}\)\Theta^W (t-t^\prime).\L{22}
\EE
Here $J_0 \(\sqrt{ z t}\)$ is Bessel function of the first kind and zero order, and window-function $\Theta^W (t )$ is different from zero (equal to
$1$) in the time interval $0 \leq t \leq T_W$.

The operator $\h{v}_+$ in the right side of the Eq. (\ref{21}) determines the contribution of all kinds of vacuum channels to the spin coherence. The
solution (\ref{21}) is formed not only by the input signal pulse $\hat a_{in}(t)$ (the first term on the right) but also due to the initial vacuum
excitations of spin waves. In quantum theory we have to consider these processes that form the explicit expression of the operator $\h{v}_+$.
However, strictly speaking, we do not need to know this expression explicitly, because vacuum channels do not introduce any contributions in the
normal ordering averages of the operators. Of course, for physical analysis, we have to calculate not only the normal ordering, but also ordinary
averages of the operators. However, they can be easily expressed via the normal ordering values.

In the formulas (\ref{21}), (\ref{22}) and everywhere after we use dimensionless coordinates and time, which are introduced according to the
relations
\BE
\Omega t\to t ,\qquad{2g^2N}z/{\Omega}\to z .\L{23}
\EE
Respectively, there are dimensionless times of writing and reading of the signal pulses and the dimensionless length of the resonant medium
\BE
\Omega T_W\to T_W,\qquad \Omega T_R\to T_R,\qquad 2g^2NL/{\Omega}\to L.\L{24}
\EE
As for the operator $\hat b_-$, according to (\ref{20}) corresponding spin wave keeps its vacuum state during all the time:
\BE
\hat{b}_-({z}, {t}) =  \hat{b}_-( {z},0 ). \L{bm}
\EE

In further discussion apart from formulas for  $\hat b_\pm$ we will need the explicit expressions for initial spin amplitudes $\hat b_1$ and $\hat
b_2$. Since
\BE
\hat{b}_1=(\hat{b}_++\hat{b}_-)/\sqrt2,\qquad\hat{b}_2=(\hat{b}_+-\hat{b}_-)/\sqrt2, \L{b1b2}
\EE
then it is easy to get that at the end of the writing process $t=T_W$ amplitudes of spin waves are given by the expressions
\BEA
&& \hat{b}_{1,2}(z)= - \frac{1}{2} \int^{T_W\!\!\!\!\!\!}_0 dt \; \h{  a}_{in}(  T_W -  t) \;G_{ab}( z, t)
+\hat v_{1,2}( z, T_W),\qquad \hat{b}_{1,2}(z) \equiv \hat{b}_{1,2}(z,T_W). \L{26}
\EEA
The meaning of operators $\hat v_{1,2}$ remains the same, namely, they correspond to the contributions of the various subsystems which are initially
in a vacuum state. We imply the average over these states for the resulting values.


\section{Input signal pulse\L{IV}}
Equations (\ref {26}) allow us to describe the quantum-statistical properties of spin waves (the quantum state of the atomic ensemble) with the help
of known properties of the input signal pulse. The latter, for example, can be postulated formally by defining various moments on the input edge of
the cell memory. However, we believe that more consistent is to define a particular source of light, the quantum statistical properties of which are
well described theoretically. For example, it could be a synchronized sub-Poissonian laser or optical-parametric generator. Statistical properties of
these sources were described in detail in articles \cite{Golubeva-2008,Golubev-2009}.

In what follows the sub-Poissonian laser will be discussed as the source of light carried quantum information to store it in the cell. Choosing phase
conditions, we can achieve that $X$-quadrature is squeezed and normally ordered correlation function can be written for its fluctuations
\cite{Golubeva-2008,Golubev-2009}:
\BE
\langle :  \delta \hat X_{in}( t) \delta \hat X_{in}({t}^\prime)  : \rangle = \frac{p}{8} \frac{\kappa(1
- \mu)}{1 - \mu/2}\; e^{\ds- \kappa (1 - \mu/2) | {t} -
{t}^\prime|}\;\Theta^{W}(t)\Theta^{W}(t^\prime).\L{28}
\EE
Let us recall the notations in this formula. Hermitian quadratures $\hat X_{in}(t)$ and $\hat Y_{in}(t)$ are introduced in the usual way as the real
and imaginary parts of Heisenberg amplitude:
\BE
\hat a_{in}(t)=\hat X_{in}(t)+i\hat Y_{in}(t).
\EE
The value $\kappa$ is a spectral width of laser mode; the parameter $p$ determines a degree of ordering of the excitation of the laser medium
$-1<p<\infty$: $p=0$ corresponds to the quite random Poissonian statistics; when $p<0$ the sub-Poissonian statistics takes place, moreover, when
$p=-1$ there is a strictly regular pumping; $p>0$ corresponds to the super-Poissonian statistics. Synchronization of the laser is ensured by the
external week field in the coherent state. Quantitatively it is described by the parameter $\mu$, equal to the ratio of the power of this field
inside the cavity to the power of generation. We choose $\mu\ll 1$ to ensure the safety of the quantum properties of generation under synchronization
by the coherent external field.

The formula (\ref{28}) is preserved if we turn to the description in terms of the dimensionless time (\ref{23}). Herewith values $\kappa$ and
$\delta\hat X_{in}$ also become to be dimensionless according to $\kappa\to\kappa/\Omega$ and  $\delta\hat X_{in}/\sqrt\Omega$.

Factors $\Theta^W(t)$ on the right of (\ref{28}) are defined as above in (\ref{22}). They convert the stationary solution into the pulse one.

It is well known that light squeezing occurs when the normal ordering correlation function (\ref{28}) is negative. As one can see, it is possible
only when pumping of the laser medium is sub-Poissonian: $p<0$.

Let us rewrite (\ref{28}) in Fourier domain performing a Fourier transform in the form:
\BE
\delta \hat X_{in,\omega}=\frac{1}{\sqrt{2\pi}}\int^{+\infty}_{-\infty} dt\;\delta \hat X_{in}( t)e^{\ds
i\omega t}.
\EE
A similar expression holds for the second quadrature $\delta \hat Y_{in,\omega}$. One can obtain a nonzero commutation relation for the spectral
components:
\BE
\[\delta \hat X_{in,\omega},\delta \hat Y_{in,\omega^\prime}\]=i/2\;\delta(\omega+\omega^\prime).\L{40.}
\EE
In further calculations, it is convenient to turn the continuous frequency scale into a discrete one. The "grain"\ of this scale we determine as
$2\pi/T_W$, then the whole set of frequencies is given by
\BE \omega \to\omega_n=({2\pi/T_W})\;n, \qquad \omega^\prime\to\omega_m = ({2\pi/T_W})\;m, \qquad n,\;m = 0,\pm1,\pm2,\ldots \EE
Now we can do the following substitutions in the Eq. (\ref{40.}): transform the delta function into the Kronecker symbol in accordance with
$2\pi/T_W\:\delta(\omega+\omega^\prime)\to \delta_{\omega_n,\:-\omega_m}$. Then, to preserve the canonical form of the uncertainty relation, we
renormalize the spectral quadratures in accordance with $\sqrt{2\pi/T_W}\:\delta \hat X_{in,\omega}\to \delta \hat X_{in,\omega_n}$ and
$\sqrt{2\pi/T_W}\:\delta \hat Y_{in,\omega^\prime}\to \delta \hat Y_{in,\omega_m}$. As a result, instead of (\ref{40.}) we obtain
\BE
\[\delta \hat X_{in,\omega_n},\delta \hat Y_{in,\omega_{m}}\]=i/2\;\delta_{\omega_n,\:-\omega_{m}}.\L{40}
\EE
We should note that the discretization of the frequency scale is associated with the ability to find appropriate grain for this. Such a grain is not
always obviously determined but in our case we can well justify the choice made above. Indeed, according to our requirements the distance between
adjacent spectral components in the discretized scale is $2\pi/T_W$. This value is much smaller than the spectral width $\kappa$ that allows us to
follow the specific spectral behaviour of the system with a good accuracy.

Now we can get from (\ref{28}):
\BE
\langle : \delta \hat X_{in,\omega_n}\; \delta \hat X_{in,\omega_{m}}  : \rangle =
 \frac{p}{4}\;\frac{\kappa^2(1-\mu)}{\kappa^2(1-\mu/2)^2+\omega^2_{n} }\;\delta_{\omega_n,\:-\omega_{m}}.\L{30.}
\EE
Hereafter we will omit the indices $n,\;m$ implying them everywhere.

The squeezing can be characterized by the quantity $\langle |\delta \hat X_{in,\omega}|^2 \rangle$ given in the form:
\BE
4 \langle |\delta \hat X_{in,\omega}|^2\rangle =1+4 \langle: |\delta \hat X_{in,\omega}|^2:\rangle=1 +
\frac{p\kappa^2(1-\mu)}{\kappa^2(1-\mu/2)^2+\omega^2}.\L{33}
\EE
One can see, when $p = -1$ (the regular laser pumping) the maximum possible two-frequency quadrature squeezing takes place, so that at zero frequency
\BE
4 \, \langle |\delta \hat X_{in,\omega=0}|^2\rangle=\mu^2/4\ll 1. \L{34}
\EE
%


\section{Squeezing of quadratures of spin waves\L{V}}
Let us now consider the signal pulse in squeezed state according to (\ref{33}) when $p<0$ at the input of the memory cell. Spin waves of the resonant
medium are formed due to the absorption of the signal pulse, and described by Heisenberg amplitudes:
\BE
\hat{b}_j(z) =\hat{X}_j(z)+i\hat Y_j(z),\qquad j=1,2.
\EE
It is interesting to find out how squeezing of the signal maps on the medium. Let us follow the quadrature components of the spin waves after the
writing process. We are interested only in $X$-quadratures, since they are directly dependent on the squeezed quadrature of the signal. According to
(\ref{26}), normal ordering averages of spin quadratures are derived in the form:
\BY
&&\< : \delta\hat X_j(z) \delta\hat X_j(z^\prime) : \> = \frac14 \int^{ T_W\!\!\!\!\!\!}_0 \int^{
T_W\!\!\!\!\!\!}_0 dt dt^\prime \; \< :\delta\hat X_{in}( T_W -
t)\delta\hat X_{in}( T_W - t^\prime) : \> G_{ab}( z, t)G_{ab}( z^\prime, t^\prime)=\nn\\
&&= \frac{p\kappa}{32} \;\frac{1 - \mu}{1 - \mu/2}\int^{T_W}_0
dt\int_0^tdt^\prime\; e^{\ds -\kappa(1-\mu/2)(t-t^\prime)} G_{ab}( z, t)G_{ab}( z^\prime,
t^\prime)+\{z\rightleftarrows z^\prime\}. \L{35}
\EY
The second equality is written taking into account Eq. (\ref{28}). Here the exponential function can be replaced by $\delta$-function under the
condition $\kappa T_W\gg1$:
\BE
\kappa (1 - \mu/2) \;e^{\ds -\kappa(1-\mu/2)(t-t^\prime)}\to\delta(t-t^\prime).
\EE

We rewrite this equation in the Fourier domain according to transformation
\BE \delta\hat X_{j,k}=\frac{1}{\sqrt{ 2\pi}} \int_{-\infty}^{+\infty} dz \;\delta\hat X_j(z)e^{\ds-ikz}\,.\L{36} \EE
Then we discretize scale $k$ with the "grain"\ $2\pi/L$ by analogy with derivation of the Eq. (\ref{33}). As a result we get
\BE
4\< |\delta\hat X_{j,k} |^2 \> = {1} + \frac{p }{4}\; \frac{(1 - \mu)}{(1 - \mu/2)^2}\int^{ T_W}_0 \!  d
t\; | G_{ab}(k, t)|^2 ,\L{37}
\EE
where $G_{ab}(k, t)$ and $G_{ab}(z, t)$ are related to each other as
\BE
G_{ab}(k, t)=\frac{1}{\sqrt L}\int_0^Ldz\;G_{ab}(z, t)e^{\ds -ikz}.\L{38}
\EE
As one can see from (\ref{37}), when $p<0$ quadratures of spin waves are squeezed. Let us remind that in the case when $p=-1$ the initial pulse was
in the perfectly squeezed state. Its excitation has been randomly distributed between two spin waves. The behaviour of quantum oscillators here is
similar to the light passed through the symmetrical beamsplitter, therefore we can expect that the squeezing of each of spin waves can not exceed
50\%. Below we demonstrate this numerically.


\section{Entanglement of spin waves\L{VI}}
We apply the Duan criterion (see Appendix A) to estimate the degree of the entanglement of two quantum subsystems. To formulate this correctly one
should specify the canonical pair (generalized  coordinate and momentum). It is easy to do on the basis of spectral amplitudes $\hat b_{i,k}$ (i=1,2)
(\ref{36}) separating their real and imaginary parts:
\BE
\hat b_{1,k}=\hat Q_{1,k}+i\hat P_{1,k},\qquad\hat b_{2,k}=\hat Q_{2,k}+i\hat P_{2,k}
\EE
\BE
\[\hat Q_{i,k},\hat P_{j,k^\prime}\]=i/2\;\delta_{ij}\;\delta(k-k^\prime)
\EE
Operators $\h{Q}_{i,k}$ and $\h{P}_{i,k}$ can play the role of the canonical variables since they are Hermitian and obey the canonical commutation
relations.

We can consider the issue of the entanglement of any two oscillators with any wave numbers $k$ and $k^\prime$, which belong to different spin waves
(or even the same wave). However, here we consider only the case of two oscillators from different spin waves with wave numbers $k$ and
$k^\prime=-k$. For this case, the Duan-criterion can be written in the form of inequality:
\BE
D_{k,-k}=\langle(\delta\hat Q_{1,k}+\delta\hat Q_{2,-k})^2\rangle+\langle(\delta\hat P_{1,k}-\delta\hat
P_{2,-k})^2\rangle<1.\L{40}
\EE
The positive value $D_{k, -k}$ is often called as the spectral parameter of the entanglement.

According to (\ref{A8})-(\ref{A9}), canonical operators $\hat Q_{i,k}$ and $\hat P_{i,k}$ can be rewritten in terms of non-Hermitian quadrature
spectral components $\hat X_{i,k}$ and $\hat Y_{i,k}$. Then the Duan-criterion (\ref{40}) will be as follows:
\BE
0<D_{k,-k}=\langle|\delta\hat X_{1,k}+\delta\hat X_{2,k}|^2\rangle+\langle|\delta\hat Y_{1,k}-\delta\hat
Y_{2,k}|^2\rangle<1\L{41}
\EE
When we derived this expression, we required that in accordance with (\ref{26}) $X-$ and $Y-$quadratures are statistically independent. This fact is
satisfied since the kernel $G_{ab}$ is real. Of course, the connection between quadratures could occur due to the properties of the signal source.
However we consider a theoretical model of laser, where the statistical correlations between quadratures are absent.

Let us rewrite inequality (\ref{41}) in terms of the normal ordering averages of operators:
\BE
-1<\langle:|\delta\hat X_{1,k}+\delta\hat X_{2,k}|^2:\rangle +\langle:|\delta\hat Y_{1,k}-\delta\hat
Y_{2,k}|^2:\rangle <0\L{42},
\EE
Due to the normal order we can ignore the vacuum contributions and in accordance with (\ref{26}) equate operators $\delta\hat X_{1,k}=\delta\hat
X_{2,k}$ and $\delta\hat Y_{1,k}=\delta\hat Y_{2,k}$. Then, returning back to the inequality (\ref{41}) for ordinary averages, we get
\BE
D_{k,-k}=4\langle|\delta\hat X_{j,k}|^2\rangle<1.
\EE
As one can see, spin waves are entangled in the same extent in which spectral quadratures of each spin wave are squeezed.


\section{Optimal observation of squeezed spin quadratures\L{VII}}
Quantum properties such as squeezing or entanglement of light (or spins) can be discussed either in spectral representation (temporal or spatial) or
in any other mode representation depending on the procedure of measurement. In the previous sections we studied the fluctuations of the spectral
quadratures $\hat X_{in,\;\omega}$ of the input light and $\hat X_{i,\;k}$ of the $i$-th $(i=1,2)$ spin wave. Alternatively, one could use, for
example, a complete orthonormal set of eigenfunctions of the memory (Schmidt modes).

Let us analyze the situation when we follow the spectral components of quadratures. According to (\ref{33})-(\ref{34}) the signal pulse at the input
of the memory cell, which was formed from the stationary radiation of the synchronized sub-Poissonian laser ($p = -1$), is in a multi-frequency
squeezed state. Herewith the maximum squeezing of the quadrature $4\langle |\delta \hat X_{in,\;\omega}|^2\rangle \ll1$ achieves almost 100\% at
$\omega=0$. After the writing of the signal pulse on the tripod atomic ensemble, two spin waves $\hat b_{1}$ and $ \hat b_{2}$ (or $ \hat b_{\pm}$)
arise instead of one wave $\h{a}_{in}$.

To estimate the effects of the squeezing and the entanglement of spin waves, we should find numerically the value $4\langle |\delta \hat
X_{i,\;k}|^2\rangle $. For the single-mode memory protocols the light-matter interaction is formally similar to the transmission of the light through
a beamsplitter. If this conclusion remains correct also in our (multimode) case then the quantum statistics of two spin waves would be the same as
for mixing of two waves in squeezed and vacuum states on the glass plate. It is easy to see that quadratures of spin waves would be entangled and
simultaneously squeezed by 50\%. Let us discuss this conclusion on the basis of the numerical analysis of Eq. (\ref{37}).

\begin{figure}
\centering
\includegraphics[height=34mm]{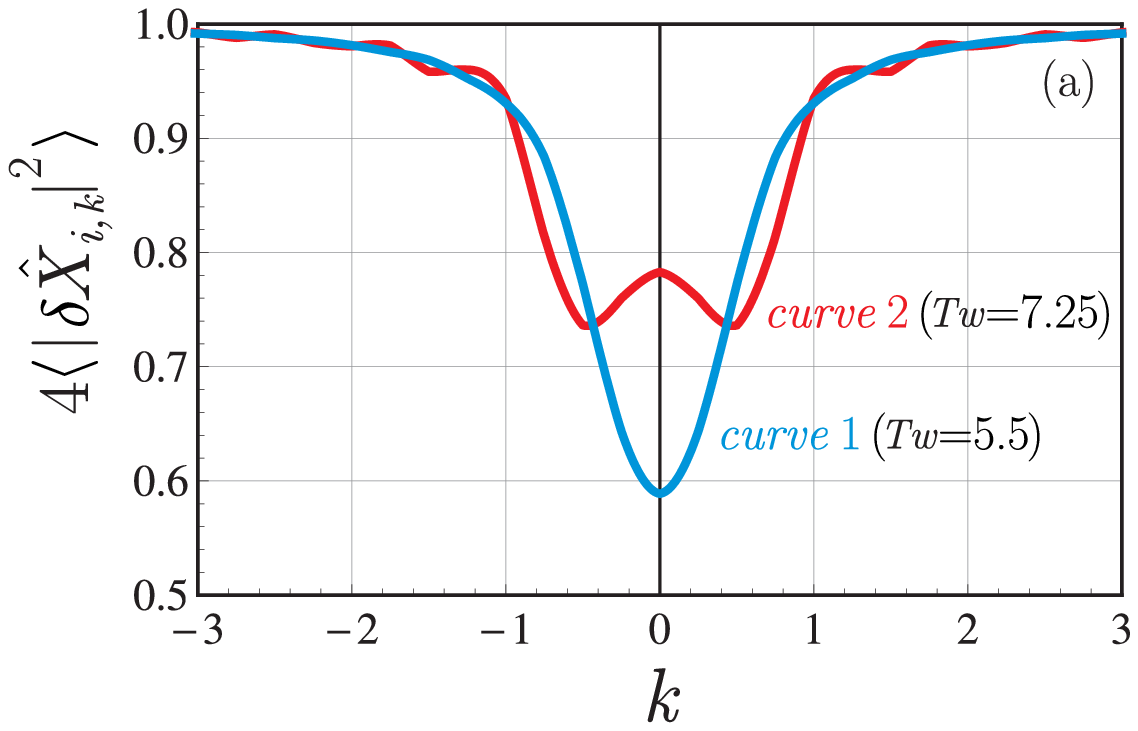}\quad\quad
\includegraphics[height=34mm]{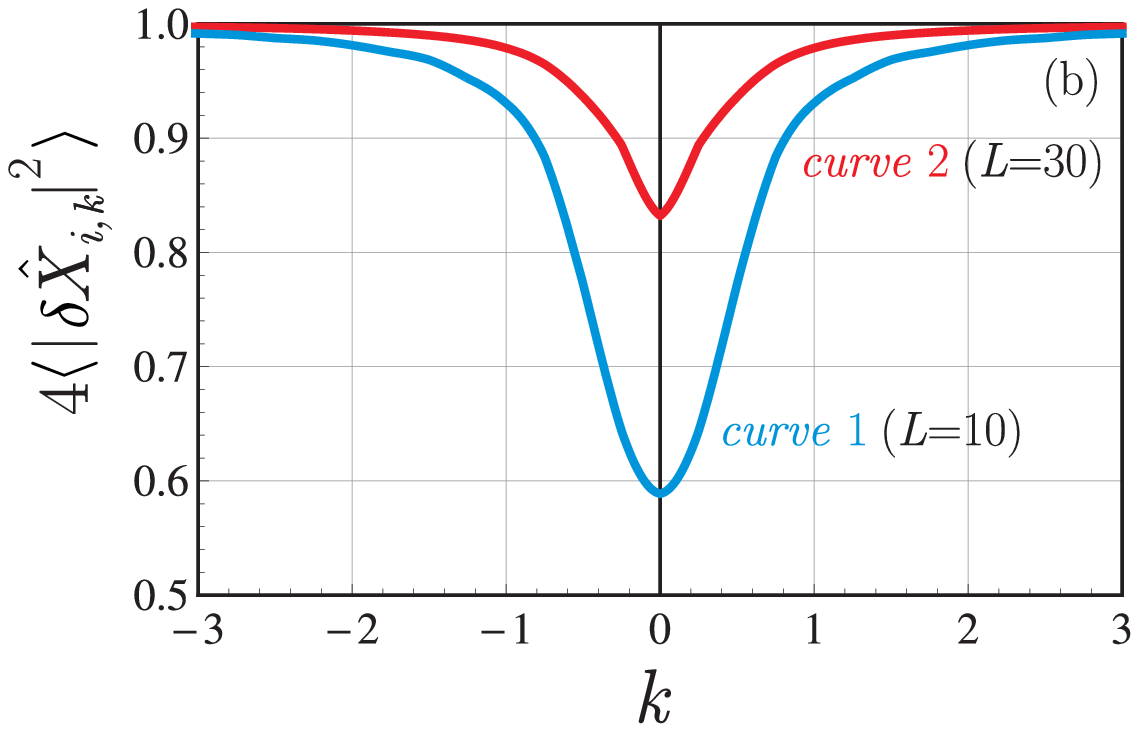}
\caption{Writing of the signal pulse in both channels: the dependence of the spectral parameter of squeezing for each spin wave $4\< |\delta\hat
X_{j,k} |^2 \>$ on the wave number $k$ with matched and unmatched sets of parameters: (a) $L = 10, T_{W} = 5.5$ (curve 1, blue), $L = 10, T_{W} =
7.25$ (curve 2, red); (b) $L = 10, T_{W} = 5.5$ (curve 1, blue),  $L = 30, T_{W} = 5.5$ (curve 2, red).} \L{pict23}
\end{figure}
\begin{figure}
\centering
\includegraphics[height=34mm]{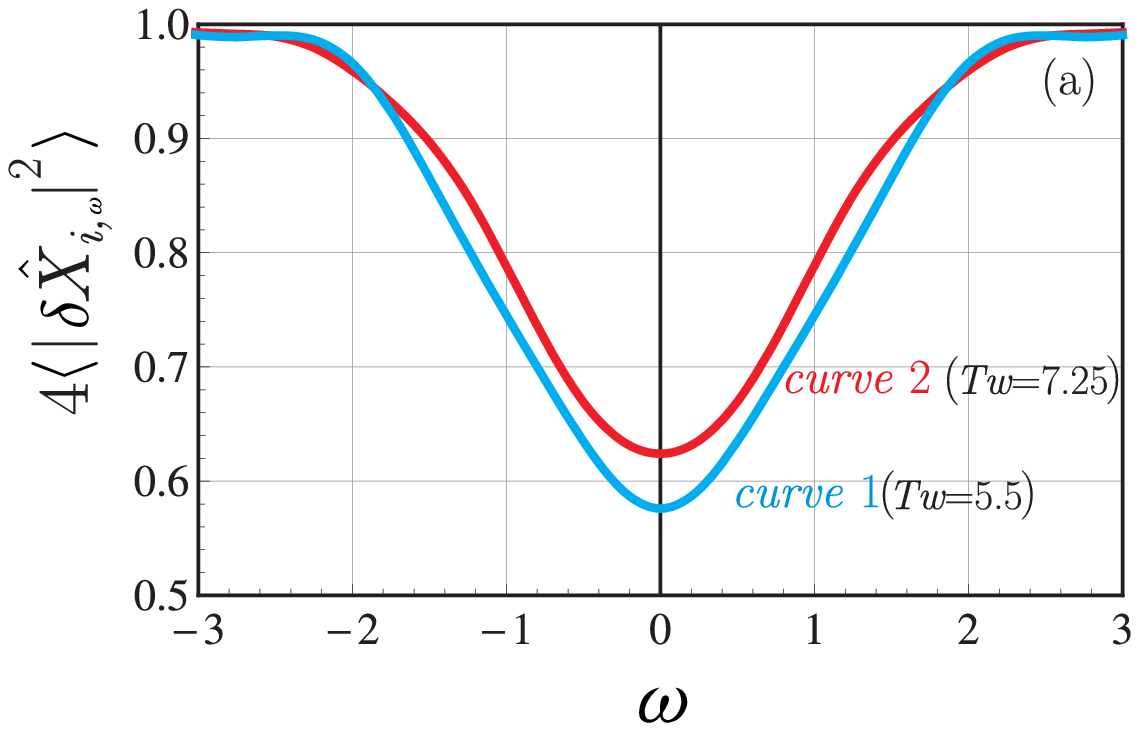} \quad\quad
\includegraphics[height=34mm]{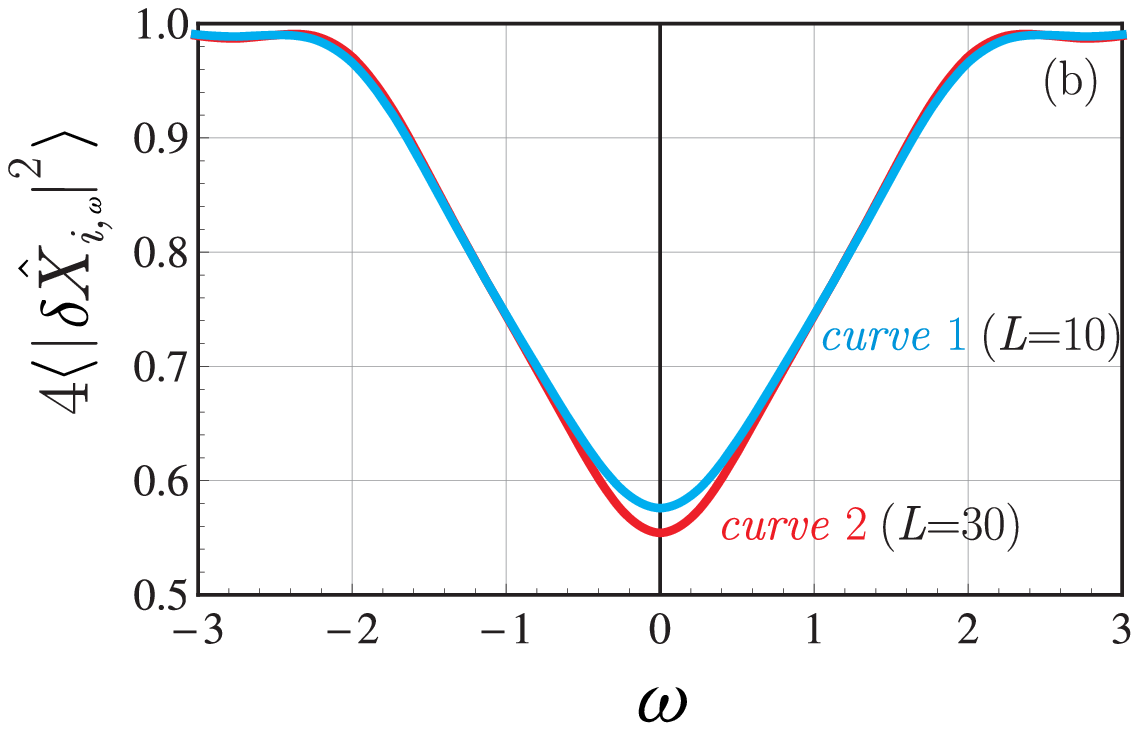}
\caption{Reading out of the signal pulse from one channel: the dependence of the spectral parameter of squeezing $4\< |\delta\hat X_{out,\omega} |^2
\>$ on the frequency $\omega$ with matched and unmatched sets of parameters: (a) $L = 10, T_{W} = 5.5$ (curve 1, blue), $L = 10, T_{W} = 7.25$ (curve
2, red); (b) $L = 10, T_{W} = 5.5$ (curve 1, blue), $L = 30, T_{W} = 5.5$ (curve 2, red).} \L{pict45}
\end{figure}
In our case of the high-speed resonant memory, losses of the signal are determined by two factors. One of them is associated with photon leakage,
when a part of signal photons are not absorbed during the writing process. Another factor is associated with a population of the upper atomic level.
These losses can be minimized by matching of two parameters: the dimensionless thickness $L$ of the memory cell and the dimensionless time $T_W$ of
the writing process. The set of parameters which provides minimum losses for given optical depth we will refer as "matched"\ or "well-matched"\ set.
For example, for $L = 10$ the requirement of minimum losses satisfied at $T_W = 5.5$ \cite{Golubeva-2011,Golubeva-2012}. Curve 1 in
Figs.~\ref{pict23}a and \ref{pict23}b corresponds to this choice. As one can see, the squeezing (and the entanglement) of spin waves reaches almost
50\% at zero wave number $k = 0$. (The difference from 50\% is determined by the choice of the relatively small value of $L$, achievable in
experiments.) Thus, in this case the analogy with the mixing of squeezed and vacuum light on the beamsplitter is quite justified.

Now let us slightly increase the dimensionless time of the signal writing $T_W$, keeping the same dimensionless thickness of layer $L$. For example,
instead of the matched set $(L=10,\; T_W=5.5)$ we choose a set $(L=10, \;T_W=7.25)$. The curve 2 in Fig.~\ref{pict23}a represents this case. One
could conclude from the comparison of the curves 1 and 2 that squeezing (and entanglement) of spins noticeably decreases in the second case. This
conclusion seems perfectly natural because we have abandoned the requirement of the minimum losses and, in general, this may lead to the destruction
of quantum properties. However, as we will show below, in this case such a conclusion is hasty and incorrect. The specified behavior of curve 2 is
caused by the nonoptimal measurement basis (that is, in the present case we follow the incorrect degree of freedom, not a degree of freedom where the
squeezing was mapped). If we want to assess more properly the quantum properties of spin waves, we should pass from the spectral $k$-components of
quadratures to the basis of eigenfunctions of the resonant medium (Schmidt modes).

The next example of unmatched set $(L=30,\;T_W=5,5)$ demonstrate this even more obvious. Here in comparison with the well-matched set, we remain
unchanged the writing time, but significantly increase the thickness of the memory cell, that, as well known, must leads to the improvement of
memory. However, in Fig.~\ref{pict23}b (curve 2) one can see that the spectral components are squeezed worse than for the well-matched set. Here we
also see the result of nonoptimal measurement procedure, but not the real deterioration of squeezing. Let us confirm this conclusion by analyzing the
value of squeezing in the retrieved light. Suppose we read out the signal only from one channel. Curve 1 in Figs.~\ref{pict45}a and \ref{pict45}b
(blue curve) shows the spectral squeezing for the matched set of parameters. Curves 2 (red curves) correspond to unmatched sets. As one can see from
the figures, the deviation from the matched set of parameters influences slightly on the squeezing of the read out signal, whereas the spectral
squeezing of spin waves changes essentially.

Thus, when we discuss the quantum state of the atomic ensemble, which arises due to the mapping of the signal, we should take into account that the
spectral basis can be unsuitable for this purpose. Alternatively, it is reasonable to use the complete orthonormal set associated with light-matter
interaction in the memory process (Schmidt modes).


\section{Full memory cycle (writing, storage and reading): Schmidt modes\L{VIII}}
In the considered model of quantum memory the squeezed signal pulse is mapped on the tripod medium so that the initial quantum correlations of light
(squeezed state) are converted into two types of correlations in the medium: quadratures each of spin waves are partially squeezed, and the waves are
entangled. Let us remind, a similar picture is observed by mixing the signal pulse in squeezed state and the vacuum field on the beamsplitter.

After the excitation of two spin waves by the single signal pulse in the medium, there are two ways for its read out. We can obtain, respectively,
one or two signal pulses at the output of the memory cell if on the reading stage we apply two control pulses simultaneously in both channels or
sequentially. In the first case, we expect that the quantum (squeezed) state of the initial pulse can be well restored. In the second case, it is
expected that the states of two pulses will copy the states of two spin waves.

Let us again solve the system of differential equations (\ref{17})-(\ref{EQbp}) with different initial conditions and represent the amplitude of the
signal field at the output of the memory cell in the form:
\BE \h{a}_{out}(t) = \h{X}_{out}(t)+i\h{Y}_{out}(t)=\int^{T_W}_0 \! dt^\prime \; \h{a}_{in}(T_W-t^\prime) G(t,t^\prime) + \h{v}(t).\L{44} \EE
Here the operator $\h{v}$ again shows the presence of vacuum channels in the system. The explicit form of the kernel $G(t,t^\prime)$ depends on the
selected geometry of the memory: whether we consider forward read out (in the same direction in which the writing was carried out) or backward one
(in the opposite direction). In accordance with \cite{Gorshkov-2007, Golubeva_Giacobino-2011}, the backward read out is more effective, when one
could neglect diffraction of light. We must note, that diffraction can significantly distort the picture. However, let us remind that in this article
we have built the theory in one-dimensional approximation, that is, we have assumed in advance that diffraction can be neglected. For the backward
reading \cite{Tikhonov-2014}:
\BE G(t,t^\prime) =\frac{1}{2}\int_0^Ldz\;G_{ab}(z,t)G_{ba}(z,t^\prime).\L{Gtt} \EE
Here the kernel $G_{ab}$ acts on the writing stage (see (\ref{22})), and  $G_{ba}$ - on the retrieving stage. Both kernels are real and in our case
they are coincide. The kernel $G(t,t^\prime)$ is Hermitian (it is real and symmetric relative to permutation of arguments). This is evident when
$T_W=T_R$, and here for the sake of simplicity, we restrict ourselves to this case. Due to the Hermitian character of the kernel, one can derive the
equation for its eigenfunctions and eigenvalues in the form:
\BE
\sqrt\lambda_i\;\varphi_i(t)= \int_0^{T_W\!\!\!} dt^\prime \varphi_i(t)\;G(t,t^\prime) .\L{46}
\EE
The solutions of this equation form a complete orthonormal set:
\BE
\int_0^{T_W\!\!\!} dt \; \varphi_i(t)\varphi_j(t)=\delta_{ij},\qquad
\sum_i\varphi_i(t)\varphi_i(t^\prime)=\delta(t-t^\prime).
\EE
It is easy to see that the Eq. (\ref{46}) generates Schmidt decomposition:
\BE
G(t,t^\prime)=\sum_i\sqrt{\lambda_i}\varphi_i(t)\varphi_i(t^\prime).\L{48}
\EE
Although the kernel $G_{ab}(z,t)$, which describes only half of the process (writing or reading), is not Hermitian, one can derive a decomposition
similar to Schmidt decomposition (\ref{48}) \cite{Tikhonov-2015}:
\BE
G_{ab}(z,t)=\sum_i(4\lambda_i)^{1/4}g_i(z)\varphi_i(t),
\EE
where the set of functions
\BE g_i(z)= {(4\lambda_i)^{-1/4}}\int_0^{T_W}dt \;\varphi_i(t)G_{ab}(z,t) \EE
is also complete and orthonormal:
\BE
\int_0^{L\!\!\!} dz \; g_i(z)g_j(z)=\delta_{ij},\qquad \sum_ig_i(z)g_i(z^\prime)=\delta(z-z^\prime).
\EE
%

\section{Simultaneous reading out from two channels\L{IX}}
In this section, we will assume that both the control pulses act simultaneously during the reading time. The corresponding Rabi frequency $ \Omega_1
$ and $ \Omega_2 $, in general, can be chosen arbitrarily, but here we will assume that $\Omega_1=\Omega_2\equiv \Omega$. In this case, the
analytical equations are the most simple.

Taking into account Eq. (\ref{44}), it is easy to get the equality for the normally ordered correlation function for the retrieved light $\langle :
\delta\hat X_{ out}(t)\; \delta\hat X_{out}(t^\prime) : \rangle$ expressed through the input correlation function:
\BE
\langle :  \delta\hat X_{ out}(t)\; \delta\hat X_{out}(t^\prime) : \rangle =
\iint\limits_0^{T_W\!\!\!\!\!\!\!\!\!\!}dt_1dt_2\langle : \delta\hat X_{in}(T_W-t_1) \delta\hat
X_{in}(T_W-t_2) : \rangle G(t,t_1)G(t^\prime,t_2). \L{Xaout}
\EE
Let us apply the Schmidt decomposition (\ref{48}) for the kernels $G$ under the sign of integration as well as decompose the input quadrature:
\BE
\delta\hat X_{ in}(T_W-t)=\sum\delta\hat x_{in,i}\varphi_i(t), \qquad\hat
e_{in,i}=x_{in,i}+iy_{in,i},\qquad
\[\hat e_{in,i},\hat e_{in,j}\]=\delta_{ij}, \L{57}
\EE
where $ \h{e}_{in, i} $ --- operator coefficients of the decomposition. Then we arrive at the equation
\BE
\langle :  \delta\hat X_{ out}(t)\; \delta\hat X_{out}(t^\prime) : \rangle = \sum_{i,j}\langle :
\delta\hat x_{ in,i}\; \delta\hat x_{in,j}: \rangle\sqrt{\lambda_i\lambda_j}\varphi_i(t)\varphi_j(t^\prime).
\L{Xaout1}
\EE
To calculate the normally ordered average $\langle :  \delta\hat x_{ in,i}\; \delta\hat x_{in,j}: \rangle$ let us use inverse transformation with
respect to the decomposition (\ref{57}), which is given by
\BE
\delta\hat x_{in,i}=\int_0^{T_W}dt\;\delta\hat X_{ in}(T_W-t)\;\varphi_i(t). \L{}
\EE
Then one can derive
\BE
\langle :  \delta\hat x_{ in,i}\; \delta\hat x_{in,j}:
\rangle=\int_0^{T_W}dt\int_0^{T_W}dt^\prime\langle:\delta\hat X_{ in}(T_W-t)\delta\hat X_{
in}(T_W-t^\prime):\rangle\; \varphi_i(t)\varphi_j(t^\prime). \L{}
\EE
Now we again take into account Eq. (\ref{28}) under the condition  $\kappa T_W\gg1$, and we get
\BE
\langle :  \delta\hat x_{ in,i}\; \delta\hat
x_{in,j}:\rangle=\frac{p}{4}\frac{1-\mu}{(1-\mu/2)^2}\;\delta_{ij}.
\EE
Substituting this equality to (\ref{Xaout1}) and passing to the Fourier domain, we obtain for $\omega =-\omega^\prime$:
\BE
4 \langle |\delta\hat X_{ out,\omega}|^2\rangle = 1 + \frac{p \; (1 - \mu)}{(1 -
\mu/2)^2}\sum_i\lambda_i|\varphi_{i,\omega}|^2. \L{62}
\EE
Let us recall that this equation was calculated under the condition $T_W=T_R$, but it can be generalized to an arbitrary relation between the writing
and reading times, which we will not consider here. As before we implied the discretization of the frequency scale, that results in
\BE
\delta\hat X_{out,\omega}=\frac{1}{\sqrt{T}}\int_0^{T}dt\delta\hat X_{ out}(t)e^{\ds i\omega t}dt,\qquad
 \varphi_{i,\omega}=\frac{1}{\sqrt{T}}\int_0^{T}dt\varphi_i(t)e^{\ds i\omega t}dt.\qquad
\L{63}
\EE
\begin{figure}
\centering
\includegraphics[height=34mm]{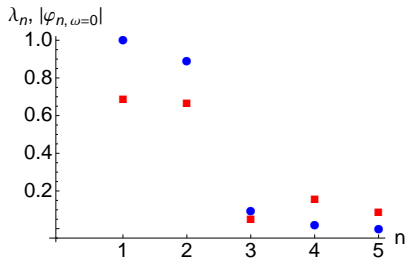}
\caption{The dependence of eigenvalues $\lambda_n$ (blue circles) and module of zero spectral components of eigenfunctions $|\varphi_{n,\omega=0}|$
(red squares) on the mode number $n$.} \L{Fig3}.
\end{figure}

Let us consider the numerical evaluations of squeezing of the spectral components of the retrieved signal after the full cycle of memory. We will use
above mentioned well-matched set of the dimensionless parameters $L = 10,\;T_W =5.5$. Under this choice of parameters only the first two modes play a
significant role in the writing/reading process as one can see from the estimation of eigenvalues $\lambda_i$ and eigenfunctions $\varphi_{i,\omega}$
(see~\ref{Fig3}). For these modes $(\lambda_1 = 1.0,\;| \varphi_{1, \omega=0}| = 0.69)$ and $(\lambda_2 = 0.9$, \;$|\varphi_{2, \omega=0}| = 0.66)$.
These parameters are negligibly small for all the other modes and their contributions in the series (\ref{62}) can be omitted.

Substituting these numerical values in Eq. (\ref{62}) under the condition $p=-1$ we get $4 \langle | \delta X_{out,\omega=0}|^2 \rangle \approx 0.13$
that means the zero spectral component of the retrieved light quadrature is squeezed by 87\%, herewith we assumed the perfect squeezing of the input
light.

We can conclude that although the initial squeezed pulse was written in the tripod medium into two channels with a random distribution of photons
between the channels, but the initial quantum features are almost completely retrieved in the output signal.

\section{Successive read out from two channels\L{X}}
In the previous section, we have studied the simultaneous reading from two spin waves and we have shown that the single signal pulse is squeezed as
far as the efficiency of the memory. Now we will discuss the situation when only one control pulse (for example, with the Rabi frequency $\Omega_1$)
acts in the time interval from 0 to $T_R$, and then the other one acts in the time interval from $T_0 (\gg T_R)$ to $T_0+T_R$. The time interval
between pulses must be chosen sufficiently large to avoid all possible transient processes associated with the first pulse before the arrival of the
second pulse.

For such a procedure of signal retrieval, it is follow from Eqs. (\ref{9})-(\ref{11}) and clear from physics, the reading is carried out first from
one channel and then from the other one. Both these processes are analogous to ones at $\Lambda$-configuration of atoms. This allows us to express
quadratures of both read out signals through the quadrature of spin waves in the form:
\BEA
&&\delta\hat X_{out}^{(1)}(t)=-\frac{1}{\sqrt2}\int_0^{L}dz\;\delta\hat X_1(z)G_{ba}(z,t)\Theta^{T}(t)+\h{v}_1(t),\L{64}\\
&&\delta\hat X_{out}^{(2)}(t)=-\frac{1}{\sqrt2}\int_0^{L}dz\;\delta\hat X_2(z)G_{ba}(z,t-T_0)\Theta^{T}(t-T_0)+\h{v}_2(t).\L{65} \EEA
It is important to note that the kernel $G_{ba}$ coincides with the kernel $G_ {ab}$ (see Eq. (\ref{22})), which we have analyzed above. This means
that for the read out of the first pulse we have chosen $\Omega_1=\Omega$ and $\Omega_2=0$, and for the second $\Omega_2=\Omega$ and $\Omega_1=0$.
Let us remind that on the writing stage we assumed $\Omega_1=\Omega_2=\Omega/\sqrt2$.

Taking into account Eqs. (\ref {64})-(\ref {65}), Schmidt decomposition and Fourier transform for the discrete frequency scale (\ref {63}), we arrive
at the expressions for the spectral squeezing in each of pulses:
\BE
4\langle |\delta\hat X^{(1)}_{out,\omega}|^2\rangle =4\langle |\delta\hat X^{(2)}_{out,\omega}|^2\rangle
=1+\frac{p}{2} \frac{(1 - \mu)}{(1 - \mu/2)^2}\sum_{i}{\lambda_i} |\varphi_{i,\omega}|^2. \L{75}
\EE

Let us estimate the value of the second term in (\ref {75}) on the basis of data shown in Fig.~\ref{Fig3}. One can be convinced that under the
previous matched set of parameters $L=10,\; T_W=5.5 $ in the case of the sub-Poissonian pumping of laser ($p=-1$) the squeezing in each of pulses
reaches 43.5\% (this value tends to 50\% with increasing the optical depth of the cell).

Now let us check the degree of the entanglement of the read out pulses. As we discussed above (see Appendix A), the Duan-criterion of entanglement
for two canonical oscillators can be given by
\BE
D_{\omega,-\omega}=\langle:   |\delta\hat X_{out,\omega }^{(1)}  + \delta\hat X_{out,\omega }^{(2)}|^2
:\rangle +
 \langle:   |\delta\hat  Y_{out,\omega }^{(1)} - \delta\hat Y_{out,\omega }^{(2)} |^2 :\rangle<0. \L{ }
\EE
In accordance with the previous statement, the second term (with a difference of quadratures) is equal zero. That is only the first term (with a sum
of quadratures) contributes into this value. As a result, we get inequality:
\BE D_{\omega,-\omega}=4\langle:   |\delta\hat X_{out,\omega }^{(1,2)}|^2:\rangle =\frac{p}{2} \frac{(1 - \mu)}{(1 - \mu/2)^2}\sum_{i}{\lambda_i}
|\varphi_{i,\omega}|^2<0. \L{68} \EE
One can see from the Eq. (\ref{68}) that similar to spin waves there exists not only the squeezing of light quadratures but also the entanglement of
signal pulses which reaches 43\% at $p=-1$ (and tends to 50\% with increasing of $L$). Thus, under the successive reading, the quantum states of the
signal pulses well repeat the quantum states of the spin waves.

\vspace{1cm}

\section{Conclusion\L{XI}}
In this work we considered the protocol of quantum memory based on tripod atomic configuration in the simplest case, when there is only one signal
pulse in squeezed state at the input of the system. In this case the writing of the signal is carried out in two channels and two spin waves are
excited in the resonance medium. It would seem that under this process there is an additional source of noise as compared with the memory, based on
the $\Lambda$-configuration of atoms, which associated with the random distribution of signal photons between spin waves. However, the simultaneous
read out from two channels (with the correct choice of the characteristic parameters for efficient operation) demonstrates the preservation of
quantum light features: the squeezing of the input signal is almost completely retrieved in the output light. The situation is similar to the mixture
of squeezed and vacuum fields on the Mach--Zehnder interferometer.

The presence of an additional degree of freedom (the second channel) in the system creates a possibility for various manipulations with the quantum
states. Particularly, under successive reading, the retrieved light reproduce the state of the spin waves, as opposed to the traditional approaches,
where the original state of the signal pulse is retrieved. This spin wave state can be different in dependence on the quantum state and the temporal
structure of the input signal.

Let us discuss qualitatively what happens when not one, but two successive signal pulses in orthogonally squeezed states are incident at the cell
input. As has been obtained above, choosing $\Omega_1=\Omega_2$ for the writing of the first pulse, we provide the excitation of spin waves $\hat
b_+$ with squeezed $X$-quadrature. Herewith the spin wave $\hat b_-$ is not affected and keeps in the vacuum state after the writing of the first
pulse. Then, the second pulse can be written in the memory with the control field $\Omega_1 = -\Omega_2$ that provides the excitation of the wave
$\hat b_- $ with squeezed $Y$-quadrature. At the same time the spin wave $\hat b_+$ keeps its $X$-quadrature in the squeezed state, which caused by
the first pulse. It leads to the entanglement of the spin waves $\hat b_{1,2}$ that can be converted into the retrieval signal pulses under
successive read out.
\\

The reported study was supported by RFBR (Grants 15-02-03656а, 16-02-00180a, 16-32-00594 and 16-32-00595).

\appendix
\section{Duan criterion of entanglement}
In our task we do not define formally some pure quantum state for the input signal pulse, but determine the source of light, which gives the field in
the mixed state. Due to this fact, the Duan criterion \cite{Duan-2000} looks like the most suitable for the analysis of the entanglement. This
criterion can be easily formulated for two oscillators, which are described by canonical variables $\hat q_i=\hat q^\dag_i$ and $\hat p_i=\hat
p_i^\dag$ ($i=1,2$). They obey the following commutation relations:
\BY &&\[\hat q_i,\hat p_j\]=i/2\;\delta_{ij},\qquad i,j= 1,2, \EY
and are introduced as a real and an imaginary parts of the annihilation operator  $\hat a_i=\hat q_i+i\hat p_i$.

The Duan criterion is introduced as follows. If the fluctuations of canonical variables $\delta\hat q_i=\hat q_i-\langle\hat q_i\rangle$ and
$\delta\hat p_i=\hat p_i-\langle\hat p_i\rangle$ obey the inequality
\BY
&& D=\langle\(\delta\hat q_1+\delta\hat q_2\)^2\rangle+\langle\(\delta\hat p_1-\delta\hat p_2\)^2\rangle<1,\L{29}
\EY
then these two oscillators are in the entangled state.

Let us now apply this for the electromagnetic field. We consider two beams, which propagate along the $z$-axis and can be formally described by two
Heisenberg amplitudes $\hat E_i(t)$ ($i=1,2$) in some cross-section (for example, on the output edge of some device). Balanced homodyne detection
allows us to select the quadrature components of the fields as its real and imaginary parts:
\BY
&&\hat E_i(t)=\hat X_i(t)+i\hat Y_i(t),\qquad \hat X^\dag_i(t)=\hat X_i(t),\qquad \hat Y^\dag_i(t)=\hat Y_i(t).\L{30}
\EY
Next we pass to the Fourier domain and show how the Duan criterion can be formulated in terms of spectral fluctuations of quadrature components
$\delta\hat X_{i,\omega}$ and $\delta\hat Y_{i,\omega}$. Thereby we demonstrate how to measure the entanglement in the scheme of balanced homodyne
detection of spectral field components.

Let us rewrite Eqs. (\ref{30}) in the Fourier domain:
\BY
&&\hat E_{i,\omega}=\hat X_{i,\omega}+i\hat Y_{i,\omega},\qquad \hat X^\dag_{i,\omega}=\hat X_{i,-\omega},\qquad \hat
Y^\dag_{i,\omega}=\hat Y_{i,-\omega}.
\EY
Non-Hermitian amplitudes $\hat E_{i,\omega}$ describe the behavior of the field oscillators with frequency $\omega_0+\omega$ and physically are
similar to the amplitudes $\hat a_i$ introduced in the beginning of this section. However, the spectral quadratures $\hat X_{i,\omega},\; \hat
Y_{i,\omega}$ are not canonical variables, because they are not Hermitian as well as the amplitudes $\hat E_{i,\omega}$.  By analogy with the
previous construction, let us introduce the real and imaginary parts of the amplitudes $\hat E_{i,\omega}$:
\BY
&&\hat E_{i,\omega}=\hat Q_{i,\omega}+i\hat P_{i,\omega},\qquad \hat Q^\dag_{i,\omega}=\hat Q_{i,\omega},\qquad \hat
P^\dag_{i,\omega}=\hat P_{i,\omega}.
\EY
Now the operators $\hat Q_{i,\omega}$ and $\hat P_{i,\omega}$ can play the role of canonical variables, since they are Hermitian and obey the
commutation relation
\BY
&&\[\hat Q_{i,\omega},\hat P_{i,\omega^\prime}\]=i/2 \;\delta(\omega-\omega^\prime).
\EY
First, we estimate the degree of entanglement of two oscillators with the same frequency, but from different light beams. In this case the Duan
criterion can be written in the form
\BY
&& D_1=\langle(\delta\hat Q_{1,\omega}+\delta\hat Q_{2,\omega})^2\rangle+\langle(\delta\hat P_{1,\omega}-\delta\hat
P_{2,\omega})^2\rangle<1.\L{A7}
\EY
The measuring procedure can be constructed so that $\hat Q$ and $\hat P$ are observables. This allows us to apply the Duan criterion in form
(\ref{34}) directly, to evaluate the entanglement of the state. However, if the experiment is based on the balanced homodyne detection, we should
rewrite this equation via the quadratures. This can be done, taking into account the following equalities:
\BY
&& \hat Q_{i,\omega}=\frac{1}{2}\(\hat X_{i,\omega}+\hat X_{i,-\omega}\)-\frac{1}{2i}\(\hat Y_{i,\omega}-\hat
Y_{i,-\omega}\),\L{A8}\\
&&\hat P_{i,\omega}=\frac{1}{2i}\(\hat X_{i,\omega}-\hat X_{i,-\omega}\)+\frac{1}{2}\(\hat Y_{i,\omega}+\hat
Y_{i,-\omega}\).\L{A9}
\EY
Then the inequality (\ref{34}) is given by
\BY
&& D_1=\langle|\delta\hat X_{1,\omega}+\delta\hat X_{2,-\omega}|^2\rangle+\langle|\delta\hat Y_{1,\omega}-\delta\hat
Y_{2,-\omega}|^2\rangle<1.\L{}
\EY
In the left side of this inequality we have omitted
\BY
&& -i\langle (\delta\hat X_{1,\omega}+\delta\hat X_{2,-\omega})(\delta\hat Y_{1,-\omega}-\delta\hat
Y_{2,\omega})\rangle+h.c.
\EY
that is correct when different quadratures are statistically independent.

Similarly, we can calculate the correlation between oscillators in different beams with arbitrary frequencies $\omega$ and $\omega^\prime$. Let us
derive Duan criterion in terms of the canonical variables for the case when $\omega^\prime=-\omega$:
\BY
&& D_2=\langle(\delta\hat Q_{1,\omega}+\delta\hat Q_{2,-\omega})^2\rangle+\langle(\delta\hat P_{1,\omega}-\delta\hat
P_{2,-\omega})^2\rangle<1.\L{}
\EY
Then, substituting Eqs. (\ref{A8})-(\ref{A9}), we obtain
\BY
&& D_2=\langle|\delta\hat X_{1,\omega}+\delta\hat X_{2,\omega}|^2\rangle+\langle|\delta\hat Y_{1,\omega}-\delta\hat
Y_{2,\omega}|^2\rangle<1.\L{}
\EY
Here we have again assumed that the different quadratures are statistically independent, and have omitted
\BY
&& -i\langle (\delta\hat X_{1,\omega}+\delta\hat X_{2,\omega})(\delta\hat Y_{1,-\omega}-\delta\hat
Y_{2,-\omega})\rangle+h.c.
\EY
Finally, let us investigate the entanglement of two field oscillators with frequencies $\omega_0\pm\omega$ in the same beam. In this case, the Duan
criterion is as follows
\BY
&& D=\langle(\delta\hat Q_{\omega}+\delta\hat Q_{-\omega})^2\rangle+\langle(\delta\hat P_{\omega}-\delta\hat
P_{-\omega})^2\rangle<1.\L{}
\EY
Substituting (\ref{A8})-(\ref{A9}), we get
\BY
&& D=4\langle|\delta\hat X_{\omega}|^2\rangle<1.\L{}
\EY
As one can see, the Duan criterion is satisfied for multimode squeezed light beam.

\end{document}